\documentclass[a4paper,twocolumn,superscriptaddress,11pt,unpublished]{quantumarticle}
\pdfoutput=1
\usepackage[utf8]{inputenc}
\usepackage[english]{babel}
\usepackage[T1]{fontenc}
\usepackage{amsmath}
\usepackage{hyperref}
\usepackage{amsfonts,latexsym,fancyhdr,graphicx,bbold,amssymb}
\usepackage{ dsfont,xcolor }

\usepackage{tikz}

\newcommand{\ket}[1]{\lvert #1 \rangle}								
\newcommand{\bra}[1]{\langle #1 \rvert}               

\newcommand*{\HKADD}[1]{\textcolor{blue}{#1}}

\begin{document}

\title{Quantum computing for accelerating cross-correlations intensive applications in signal processing}
\date{02/01/2022}
\author{Valentina Caprara Vivoli}
\affiliation{Thermo Fisher Scientific, Achtseweg Noord 5, 5651 GG Eindhoven, The Netherlands}
\email{capraravalentina@gmail.com}
\author{Yuchen Deng}
\affiliation{Thermo Fisher Scientific, Achtseweg Noord 5, 5651 GG Eindhoven, The Netherlands}
\author{Kohr Holger}
\affiliation{Thermo Fisher Scientific, Achtseweg Noord 5, 5651 GG Eindhoven, The Netherlands}
\author{Erik Franken}
\affiliation{Thermo Fisher Scientific, Achtseweg Noord 5, 5651 GG Eindhoven, The Netherlands}

\maketitle

\begin{abstract}
  Despite their importance as subfields of mathematics and engineering, signal and image processing have not received much attention in the field of quantum computation.
  Cross-correlations are instrumental to all the aforementioned fields.
  In this article we help fill this void by providing two quantum algorithms, one for computing and storing cross-correlations, and one for implementing the expectation maximization maximum likelihood algorithm.
  In addition we show that the quantum expectation maximization maximum likelihood algorithm has a quadratic speed-up compared to the classical analog.
\end{abstract}

\section{Introduction}
\hspace{.5cm}
Cross-correlations are an important tool for many signal processing
applications, in particular image processing \cite{Saxton76}. They are used in various applications, where images need to be matched, for instance in object tracking where the position of a subimage is tracked through a sequence of larger images, or for template matching, i.e. to find a certain pattern in a larger image. The method is widely applied in various fields, for instance electron microscopy, where it is used e.g. for drift correction \cite{Zhang16}, tomographic tilt series alignment \cite{Frank92}, autofocus \cite{Koster92}, and pattern recognition, like template matching \cite{Best07}. Cross-correlations are an important ingredient for many other algorithms. An important algorithm is expectation maximization maximum likelihood (EMML) \cite{Dempster77}, used for example in single particle reconstruction \cite{Sigworth98,Scheres12}.  \\
Part of the reason why cross-correlations are so widely used is that they are very fast to  compute. Indeed, the complexity of the cross-correlations calculation on $N$ elements data arrays reduces from $O(N^2)$ to $O(N \log N)$ when one uses Fast Fourier Transforms (FFT) \cite{Cooley67}. This allows for reducing the complexity of multiple algorithms.\\
There are several examples of computer algorithms that can be sped up through the use of quantum computing \cite{QuantumZoo}.
Nevertheless, it is believed that cross-correlations cannot be calculated through quantum computing. Scientists have focused on trying to "compute the convolution and correlation of two quantum states" \cite{Lomont03} replacing FFTs with Quantum Fourier Transforms (QFT) \cite{Coppersmith94}. This is not possible because of the laws of quantum mechanics \cite{Lomont03}. A field in quantum computing that might be related to the computation of cross-correlations is quantum linear algebra \cite{Chakraborty19,Gilyen19,Chao20}. However, linear algebra can only be used for the computation of cross-correlations and not for the non-linear optimizations involved in almost all signal processing applications. For such reason quantum signal processing has not been developed so far in an extensive way \cite{Yan16}, even though more recent work is addressing the related field of ``quantum image understanding'' \cite{ Yan20a, Yan20b, Yan20c}\footnote{Note that these articles were published after the first release of this manuscript on the ArXiv. However, they do not cover the problem addressed in this article. }.\\
In this article we adopt a new approach to the problem of calculating cross-correlation functions through quantum computing. Instead of trying to use QFTs to calculate the cross-correlations of two quantum states, we use another quantum primitive, namely Quantum Amplitude Estimation (QAE) \cite{Brassard02}, to get the cross-correlations as an output encoded in a string of qubits. QAE is an algorithm that allows, having a state $\ket{\Psi}$, to estimate the probability that the system lies in a subspace fulfilling a list of conditions that can be encoded in a Boolean function. QAE is mainly used to solve counting problems \cite{Terhal98,Iwama10}.
The calculation of the correlations is thus made possible thanks to encoding the data as the probability of the terms that compose the initial state and not in the amplitude as attempted before.
This technique is successful only in the case of real positive data. However, we will discuss in the concluding remarks how to relax this limitation.
Furthermore, we show that the algorithm can be successfully used also for a specific signal processing application where cross-correlation functions are used, namely EMML, getting a speed-up compared to any known classical algorithm.
As a final remark, we stress the fact that such algorithm is suitable for the calculation of the convolution as well.
Such results show that, regardless of whether improvements can be made, quantum computing has the potential for speeding up signal processing.

\section{Classical algorithms}
\hspace{.5cm}
In this section we provide some background on classical algorithms for which we present a quantum alternative in the following paragraphs.

\subsection{Discrete Cross-correlation function}
\hspace{.5cm}
The expression for discrete cross-correlation functions between two N-complex valued arrays $\left\{x_i\right\}$ and $\left\{y_i\right\}$ is
\begin{equation}
C_j=\sum_{i=0}^{N-1}x_i^*y_{j\oplus i},\label{eq:CorssCorrelation}
\end{equation}
where x and y are discrete (sampled) 1D signals, $*$ denotes the complex conjugate, and $\oplus$ represents the cyclic sum mod N. The latter is used since we are assuming periodic signals. If this is not desired in the application, precautions must be taken, e.g. by padding the signal. The operation could be interpreted as follows: a discrete signal $y$ is periodically shifted by $j$ and subsequently, for each $j$ in $\{0,...,N-1\}$\HKADD{,} an inner product is computed between shifted $y$ and non-shifted $x$, which is the resulting cross-correlation $C_j$.
It is possible to calculate cross-correlation functions using FFTs \cite{Cooley67} with a complexity of $O(N\log N)$.\\
Discrete cross-correlations can also be defined when the arrays have two different sizes, $N$ and $M$ respectively, with $M<N$. In this case the complexity is $O(MN)$, when cross-correlations are calculated in a straightforward way, or $O(N \log N)$, when one pads the smaller array with zeros.

\subsection{EMML algorithm}
\hspace{.5cm}
  EMML is an iterative method for determining the parameters of a statistical model that contains unobserved (latent) variables, in such a way that it best fits observed data \cite{Dempster77,Sundberg74,Sundberg76}.
  If $X$ is an array of observations, $Y$ are the unknown parameters of the model and $\delta$ the latent variables of the statistical model, the best-fitting model is defined as having parameters that maximize the likelihood of seeing the data, given the model:
  \begin{align}
    \label{eq:ml_estimate}
    Y_{\mathrm{EM}} &= \text{arg}\max_Y L(Y; X), \\
    \label{Eq:Likelihood}
    L(Y; X) &:= p(X | Y) = \int p(X, \delta\, |\, Y)\, \mathrm{d}\delta.
  \end{align}
  Directly maximizing the likelihood is usually intractable.
  Therefore the maximization is carried out in two separate steps: \\[1em]
  \textbf{E Step:} Compute the expectation of the log-likelihood $\log L(Y; X, \delta) = \log p(X, Y | \delta)$ with respect to the distribution of the latent variable $\delta$, conditioned on the observations $X$ and the current estimate $Y^t$ of $Y$:
  \begin{equation}
    \label{eq:E_step}
    \begin{split}
      E(Y; Y^t)
      &= \mathds{E}_{\delta | X, Y^t}[\log L(Y; X, \delta)] \\
      &= \int p(\delta | X, Y^t)\, \log L(Y; X, \delta)\, \mathrm{d}\delta.
    \end{split}
  \end{equation}
  \textbf{M Step:} Update the parameters by maximizing $E$ with respect to $Y$:
  \begin{equation}
    \label{eq:M_step}
    Y_{t+1} = \text{arg}\max_Y E(Y; Y^t).
  \end{equation}
  In general, this optimization procedure is highly nonlinear in the inputs $X$.

  For the specific problem of matching a set of shifted images with Gaussian noise, however, the iterations can be simplified.
  Let us assume that we observe a set of noisy images $X^{(i)},\ i=1,\dots,N$, subject to unknown shifts $s_i$.
  Each pixel value $X_k^{(i)}$ is i.i.d. with
  \begin{equation*}
    X_k^{(i)} \sim N(Y_{k \oplus s_i}, \sigma^2),
  \end{equation*}
  where the noise variance $\sigma^2$ is assumed to be known.
  We intend to find the noise-free unshifted image $Y$.
  The data distribution is (up to a constant factor) given by
  \begin{equation*}
    P[X_k^{(i)} = x_k^{(i)} | Y = \mathbf{y},\ \boldsymbol{\delta} = \mathbf{s}] = \mathrm{e}^{-\frac{\left(x_{k \ominus s_i}^{(i)} - y_k\right)^2}{2 \sigma^2}},
  \end{equation*}
  or in vector form
  \begin{equation*}
    P[X^{(i)} = \mathbf{x}^{(i)} | Y = \mathbf{y},\ \boldsymbol{\delta} = \mathbf{s}] = \mathrm{e}^{-\frac{\bigl\| \mathbf{x}_{\cdot \ominus s_i}^{(i)} - \mathbf{y} \bigr\|^2}{2 \sigma^2}}.
  \end{equation*}
  We further assume a uniform prior on the distribution $\boldsymbol{\delta}$ of shifts.
  The log-likelihood function for this model is (up to an additive constant) given by
  \begin{equation*}
    -\log L(\mathbf{y}; \mathbf{x},\ \delta) = \sum_i \sum_{s_i} \mathbf{1}[\delta^{(i)} = s_i] \bigl\| \mathbf{x}_{\cdot \ominus s_i} - \mathbf{y} \bigr\|^2,
  \end{equation*}
  where $\mathbf{1}[\delta^{(i)} = s_i]$ is an indicator function taking the value 1 if $\delta^{(i)} = s_i$, and 0 otherwise.
  The conditional probability for the E step can be calculated using Bayes' theorem (exploiting independence w.r.t $i$):
  \begin{align*}
    & P[\boldsymbol{\delta} = \mathbf{s} | X = \mathbf{x},\ Y = \mathbf{y}^t]\\
    &\hspace*{10pt} = \prod_i P[\delta^{(i)} = s_i | X^{(i)} = \mathbf{x}^{(i)},\ Y = \mathbf{y}^t], \\[1ex]
    & P[\delta^{(i)} = s_i | X^{(i)} = \mathbf{x}^{(i)},\ Y = \mathbf{y}^t] \\
    &\hspace*{10pt} = \frac{P[X^{(i)} = \mathbf{x}^{(i)} | \delta^{(i)} = s_i,\ Y = \mathbf{y}^t]\, P[\delta^{(i)} = s_i]}{\sum_{s'} P[X^{(i)} = \mathbf{x}^{(i)} | \delta^{(i)} = {s'},\ Y = \mathbf{y}^t]} \\
    &\hspace*{10pt} = \frac{\mathrm{e}^{-\frac{\bigl\| \mathbf{x}_{\cdot \ominus s_i}^{(i)} - \mathbf{y} \bigr\|^2}{2\sigma^2}}}{\sum_{s'} \mathrm{e}^{-\frac{\bigl\| \mathbf{x}_{\cdot \ominus s'}^{(i)} - \mathbf{y} \bigr\|^2}{2\sigma^2}}} \\
    &\hspace*{10pt} =: w_i(s_i; \mathbf{x}^{(i)}, \mathbf{y}^t).
  \end{align*}
  This expression, also called \textit{softmax function}, peaks at the correct shift value, and the sharpness of the peak increases with decreasing noise variance $\sigma^2$.
  Hence, the shift of each image can be determined from its corresponding distribution function $w_i$.
  The conditional expectation (up to irrelevant constants) in the E step now reads as
  \begin{equation*}
    E(\mathbf{y}; \mathbf{y}^t) = - \sum_i \sum_{s_i} w_i(s_i; \mathbf{x}^{(i)}, \mathbf{y}^t)\, \bigl\| \mathbf{x}_{\cdot \ominus s_i}^{(i)} - \mathbf{y} \bigr\|^2.
  \end{equation*}
  This function is easily maximized in the M step, yielding the update relation
  \begin{equation*}
    y_j^{t+1} = \frac{1}{N} \sum_i \sum_k w_i(k; \mathbf{x}^{(i)}, \mathbf{y}^t) x^{(i)}_{j \oplus k},
  \end{equation*}
  i.e., average of the ``motion-corrected'' images
  \begin{equation*}
    \left(x_j^{(i)}\right)^t = \sum_k w_i(k; \mathbf{x}^{(i)}, \mathbf{y}^t) x^{(i)}_{j \cdot \oplus k}
  \end{equation*}
  gained by cross-correlating the weight function $w_i$ with the original image $\mathbf{x}^{(i)}$.
  Since the function $w_i$ can be expensive to compute, it is common to replace it with the cross-correlation of $\mathbf{x}^{i}$ and $\mathbf{y}^t$ -- it has the same property of peaking at the true shift $s_i$.
  The simplified iteration reads as
  \begin{align*}
    & C_k^t = \sum_l x_l^{(i)} y^t_{k \cdot \oplus l}, \\
    &\left(x_j^{(i)}\right)^t = \sum_k C_k^t x^{(i)}_{j \cdot \oplus k}, \\
    & \mathbf{y}^{t + 1} = \frac{1}{N} \sum_i \left(\mathbf{x}^{(i)}\right)^t.
  \end{align*}
  Instead of using absolute shifts, one can also update the images $\mathbf{x}^{(i)}$ in each iteration and compute new incremental shifts in each step:
  \begin{align*}
    & C_k^t = \sum_l \left(x_l^{(i)}\right)^t y^t_{k \cdot \oplus l}, \\
    &\left(x_j^{(i)}\right)^{t + 1} = \sum_k C_k^t \left(x^{(i)}_{j \cdot \oplus k}\right)^t, \\
    & \mathbf{y}^{t + 1} = \frac{1}{N} \sum_i \left(\mathbf{x}^{(i)}\right)^{t + 1}.
  \end{align*}
  However, this procedure has the disadvantage that the images $\left(x_j^{(i)}\right)^{t + 1}$ become blurrier in each iteration, thus degrading the performance of the algorithm.


\section{Quantum Cross-correlations computation}
\hspace{.5cm}
\begin{figure}[h]
\begin{center}
\includegraphics[width=8.5cm]{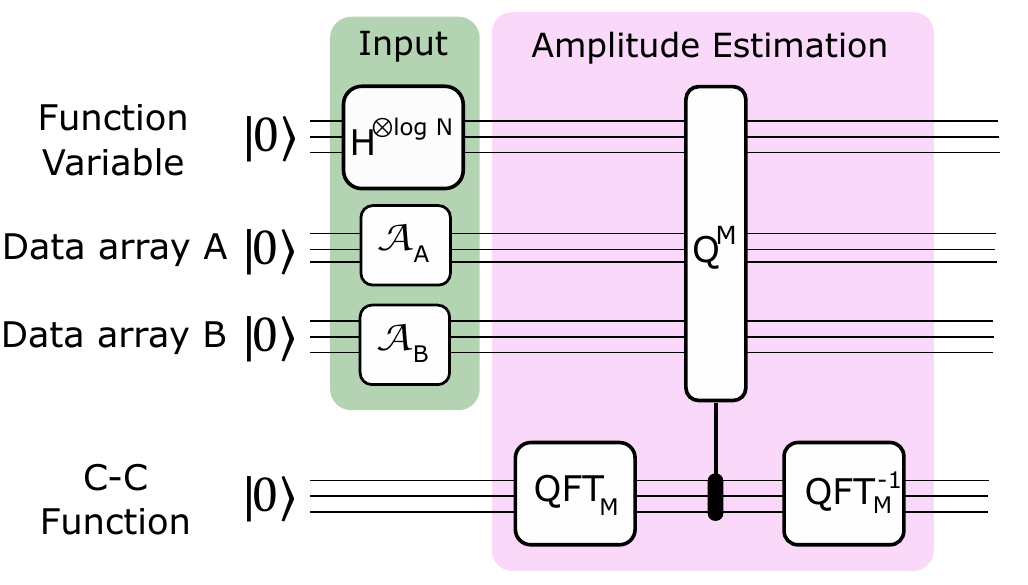}
\end{center}
\caption{Quantum circuit for the parallel calculations of the cross-correlation values. Two sets of qubits are used to store the two data arrays A and B. Another set of qubits of the same dimension is used for the superposition of all the values of the function variable. QAE is applied to the registers of the images using the shift register as an ancilla.}\label{fig:CrossCorrelation}
\end{figure}
\hspace{-0.7cm}\fbox{\parbox{\linewidth}{\textbf{Theorem 3.1. Cross-Correlation algorithm.}\\
Let $x$ and $y$ be real and positive vectors of lenght $N$ and $M$, respectively. Assume we have 3 quantum registers, each composed of $\log N$ qubits. Then there exists a quantum algorithm of complexity $O(\sqrt{N})$ that encodes in a quantum register of $\log M$ qubits the angles $\theta_j$, such that $\sin^2 \theta_j=C_j$, where $C_j$ is the cross-correlation defined in \eqref{eq:CorssCorrelation}.}}
Here we explain how to calculate the cross-correlation functions without the need of the FFTs. Consider two data arrays A and B of an even number $N$ of elements $\{x_j^A\}$ and $\{x_j^B\}$, respectively. Our goal is to achieve a protocol where we give as input the two strings A and B, and we get as output the cross-correlation elements $C_j=\sum_{\bar{\jmath}}x_{\bar{\jmath}\oplus j}^A x_{\bar{\jmath}}^B$. We do restrain our algorithm to the case where the elements $\{x_j\}$ of each string fulfill the following two constraints
\begin{enumerate}
\item $x_j\in \mathds{R}$, and $x_j\ge 0$,
\item $\sum_{j=0}^{N-1}x_j=1$.
\end{enumerate}
Note these constraints can always be achieved by transforming input data $x_j'$ linearly, i.e. $x_j=a x_j'+b$ where $a$ and $b$ are chosen such that constraints 1 and 2 are fulfilled. Such scaling does not impact the topology of the resulting cross-correlation $C_j$, and $C_j$ can be easily scaled back accordingly.
The quantum circuit is shown in Fig. \ref{fig:CrossCorrelation}.

We need 4 sets of qubits. Two are meant for the arrays A and B, respectively, both of dimension $\log N$. One set of dimension $\log N$ is meant for the variable $\bar{\jmath}$ of the cross-correlations function. The last one is of dimension $\log M$, with $M$ even integer, where the cross-correlation function is calculated. $\log M$ is the number of digits of precision we calculate the amplitude of the cross-correlations function with. We do define the following operators:
\begin{enumerate}
\item $\mathcal{A}_A$ ($\mathcal{A}_B$) as the operator that encodes arrays A (B) inside the quantum computer, i.e. $\mathcal{A}_A\ket{0}^{\otimes \log N}=\sum_{j=0}^{N-1}\sqrt{x_j^A}\ket{j}$ ($\mathcal{A}_B\ket{0}^{\otimes\log N}=\sum_{j'=0}^{N-1}\sqrt{x_{j'}^B}\ket{j'}$) and $\mathcal{A}=I_{\text{var}}\otimes\mathcal{A}_A\otimes\mathcal{A}_B$.
\item $S_0=I_{\text{var}}\otimes \left(I-2\ket{0}\bra{0}^{\otimes 2\log N}\right)_{AB}$, is the operator reversing the phase of the state $\ket{0}^{2\log N}_{AB}$, while leaving the other states unchanged.
\item $S_{j_A\ominus j'_B=\bar{\jmath}_{\text{var}}}=I_{\text{var}AB}-2 \sum_{\bar{\jmath},j=0}^{N-1}\ket{\bar{\jmath},\bar{\jmath}\oplus j,\bar{\jmath}}\bra{\bar{\jmath},\bar{\jmath}\oplus j,\bar{\jmath}}_{\text{var}AB}$ reverses the phase of the states that fulfill the condition $j_A\ominus j'_B=\bar{\jmath}_{\text{var}}$.
\item $Q=\mathcal{A} S_0 \mathcal{A}^{-1} S_{j_A\ominus j'_B=\bar{\jmath}_{\text{var}}}$, known as Grover operator.
\end{enumerate}
For simplicity we assume that the complexity of these operators is 1.
In what follows we explain the algorithm in details.

\subsection{Input initialization}
\hspace{.5cm}
The two registers A and B are initialized to store the two arrays, i.e. we have two states
\begin{equation}
\sum_{j=0}^{N-1}\sqrt{x_j^A}\ket{j}_A\text{, and }\sum_{j'=0}^{N-1}\sqrt{x_{j'}^B}\ket{j'}_B.
\end{equation}
A gate $H^{\otimes \log N}$, where $H$ is the Hadamard operator, is applied to the variable space. Therefore the total state is
\begin{equation}
\frac{1}{\sqrt{N}}\sum_{\bar{\jmath}=0}^{N-1}\ket{\bar{\jmath}}_{\text{var}}\sum_{j,j'=0}^{N-1}\sqrt{x_j^Ax_{j'}^B}\ket{j,j'}_{AB}\ket{0}^{\otimes \log M}_{\text{cor}}.
\end{equation}

\subsection{Quantum Amplitude Estimation}
The next step is a modification of what is known in literature as QAE \cite{Brassard02}.
A QFT is performed over an additional register, giving the state $\sum_{m=0}^{M-1}\ket{m}$. Then, a controlled-$Q^M$ $\Lambda_M(Q)$ is applied on the entire set of qubits. The operator $\Lambda_M(Q)$ is defined as
\begin{equation}
\Lambda_M(Q)\ket{y}_{\text{var}AB}\ket{m}_{\text{cor}}=Q^m\ket{y}_{\text{var}AB}\ket{m}_{\text{cor}}.
\end{equation}
The operator $Q$, defined above, uses the register of the variable as an ancilla that helps for defining the constraints of the QAE, but stays untouched during the other steps.
The state of the variable registers, A and B can be re-written as a function of the eigenvectors of Q, $\ket{\bar{\jmath}}\ket{\Psi_+^{\bar{\jmath}}}$, and $\ket{\bar{\jmath}}\ket{\Psi_-^{\bar{\jmath}}}$, i.e.
\begin{equation}
-\frac{i}{\sqrt{2N}}\sum_{\bar{\jmath}=0}^{N-1}\ket{\bar{\jmath}}_{\text{var}}\left(e^{i \theta_{\bar{\jmath}}}\ket{\Psi_+^{\bar{\jmath}}}_{AB}-e^{-i \theta_{\bar{\jmath}}}\ket{\Psi_-^{\bar{\jmath}}}_{AB}\right),
\end{equation}
where $\sin^2\theta_{\bar{\jmath}}=\sum_{j=0}^{N-1}x_{\bar{\jmath}\oplus j}^A x_j^B=C_{\bar{\jmath}}$ are the cross-correlation values. $\ket{\Psi_{\pm}^{\bar{\jmath}}}_{\text{AB}}$ are defined as
\begin{equation}\begin{split}
\ket{\Psi_{\pm}^{\bar{\jmath}}}=\frac{1}{\sqrt{2}}\biggl(&\frac{1}{\sqrt{C_{\bar{\jmath}}}}\sum_{j=0}^{N-1}\sqrt{x_{\bar{\jmath}\oplus j}^A x_j^B}\ket{\bar{\jmath}\oplus j,j}_{AB}\\
               &+\frac{1}{\sqrt{1-C_{\bar{\jmath}}}}\sum_{\substack{j,j'=0 \\ j'\neq j \ominus \bar{\jmath}}}^{N-1}\sqrt{x_j^Ax_{j'}^B}\ket{j,j'}_{AB}\biggl)
\end{split}\end{equation}
The eigenvalues of $\ket{\bar{\jmath}}\ket{\Psi_{\pm}^{\bar{\jmath}}}$ are $e^{\pm i 2\theta_{\bar{\jmath}}}$.
The total state after the controlled-$Q^M$ is
\begin{equation}\begin{split}
-\frac{i}{\sqrt{2N}}\sum_{\bar{\jmath}=0}^{N-1}\ket{\bar{\jmath}}_{\text{var}}\biggl(&e^{i \theta_{\bar{\jmath}}}\ket{\Psi_+^{\bar{\jmath}}}_{AB}\ket{S_{M}\left(\frac{\theta_{\bar{\jmath}}}{\pi}\right)}_{\text{cor}}\\
-&e^{-i \theta_{\bar{\jmath}}}\ket{\Psi_-^{\bar{\jmath}}}_{AB}\ket{S_{M}\left(1-\frac{\theta_{\bar{\jmath}}}{\pi}\right)}_{\text{cor}}\biggl),
\end{split}\end{equation}
where
\begin{equation}
\ket{S_{M}\left(\frac{\theta_{\bar{\jmath}}}{\pi}\right)}=\sum_{m=0}^{M-1}e^{2 i m \theta_{\bar{\jmath}}}\ket{m}.
\end{equation}
Now it is just necessary to perform an inverse QFT on the last register to get
\begin{equation}\begin{split}
-\frac{i}{\sqrt{2N}}\sum_{\bar{\jmath}=0}^{N-1}\ket{\bar{\jmath}}\bigg(&e^{i \theta_{\bar{\jmath}}}\ket{\Psi_+^{\bar{\jmath}}}\ket{M\frac{\theta_{\bar{\jmath}}}{\pi}}\\
&-e^{-i \theta_{\bar{\jmath}}}\ket{\Psi_-^{\bar{\jmath}}}\ket{M\left(1-\frac{\theta_{\bar{\jmath}}}{\pi}\right)}\bigg).
\end{split}\end{equation}
Since $\langle C_{\bar{\jmath}}\rangle=\frac{1}{N}$, it is reasonable to conclude that $\theta_{\bar{\jmath}}$ is a value of the order of $\frac{1}{\sqrt{N}}$. Hence, $M=\alpha \sqrt{N}$, where $\alpha\ll N$ can be chosen in a second moment. Thus, the complexity of the QAE is upper bounded by $O(\sqrt{N})$.
We stress here that this algorithm allows to merely store the cross-correlations inside the fourth set of qubits. If one performed a measurement of such a set, one would end up with one of the values of $M\frac{\theta_{\bar{\jmath}}}{\pi}$ or $M\left(1-\frac{\theta_{\bar{\jmath}}}{\pi}\right)$ picked at random. We summarize the above presented steps in the following algorithm.

{\bf Algorithm for calculating cross-correlation functions.}
\begin{enumerate}
\item Import the two arrays A and B;
\item Apply a $H^{\otimes \log N}$ gate to the variable $\bar{\jmath}$ register;
\item Apply a $QFT_M$ to the last register.
\item Apply the operator $\Lambda_M(Q)$.
\item Perform an inverse $QFT_M$.
\end{enumerate}

\section{Quantum Expectation-Maximization-Maximum-Likelihood algorithm}
\hspace{.5cm}
In this section we present a quantum algorithm meant for EMML.
\begin{figure}[h]
\begin{center}
\includegraphics[width=9.5cm]{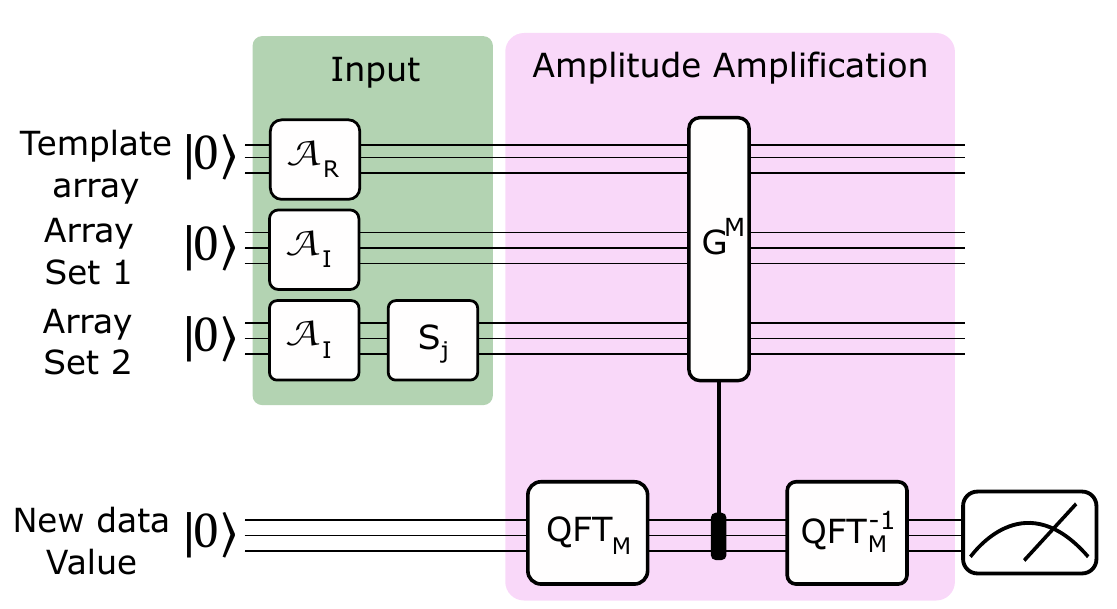}
\end{center}
\caption{Quantum circuit for the calculation of a new data value in EMML. Two sets of qubits are used to store the same data array. Another set of qubits of the same dimension is used to store the template array. QAE is applied to the registers of the arrays. Finally, a measurement on the new data register is performed.}\label{Im:EMML}
\end{figure}
\fbox{\parbox{\linewidth}{\textbf{Theorem 4.1. Quantum EMML algorithm.}\\
Suppose we have a signal of $N$ data points $\{x_{j,k}^t\}$, a template signal $\{X_{j,k}\}$ of $N$ data points and 3 quantum registers, each composed of $\log N$ qubits. Then there exist a quantum algorithm of complexity $O(\sqrt{N})$  that realizes the E and M steps of the EMML algorithm and encodes in a quantum register of $\log M$ qubits the angle $\theta_{j,k}$, such that $\sin^2 \theta_{j,k}=x_{j,k}^{t+1}$.  }}
\\Consider a set of data arrays, each one of dimension $N\times N$ where $N$ is an even number. We generate a template array, whose set of values $\{X_{j,k}^t\}$ are given by the average of the data of the set of data arrays. Here, $j$ and $k$ are the pixel coordinates and the apex $t$ refers to the iteration number. Let's focus on one specific array of elements $\{x_{j,k}^t\}$.
During the EMML at every iteration $(t+1)$ one has to calculate for each array the new data values $\{x_j^{t+1}\}$. For the sake of simplicity we focus on the case where the array transformations are limited to translations. Given an array, the equation for $x_{j,k}^{t+1}$ is
\begin{equation}
x_{j,k}^{t+1}=\sum_{\bar{\jmath},\bar{k}=0}^{N-1} C_{\bar{\jmath},\bar{k}}^tx_{j\oplus \bar{\jmath},k\oplus\bar{k}}^t,
\end{equation}
where $C_{\bar{\jmath},\bar{k}}^t =\sum_{j',k'=0}^{N-1}X_{j',k'}^t x_{j'\oplus \bar{\jmath},k'\oplus \bar{k}}^t$. \\
Now let's focus on the quantum algorithm.
For calculating the data value $x_{j,k}^{t+1}$, we need 4 registers. The first 3, the template, and the copy 1 and 2 registers, are composed of $2\log N$ qubits each and are used to store two sets of the data array and one copy of the template image. The last register is an ancillary register where the value $x_{j,k}^{t+1}$ is going to be written and is composed by $\log M$ qubits. $M$ is an even number that depends on the precision with which one wants to calculate $x_{j,k}^{t+1}$.\\
 We define the following operators:
\begin{enumerate}
\item $\mathcal{A}_I$ as the operator that encodes the data array inside the quantum computer, i.e. $\mathcal{A}_I\ket{0}^{\otimes 2 \log N}=\sum_{j,k=0}^{N-1}\sqrt{x_{j,k}^t}\ket{j,k}$.
\item $S_{j,k}$, as the gate implementing the difference operation
$$
\sum_{\tilde{\jmath},\tilde{k}=0}^{N-1}\sqrt{x_{\tilde{\jmath},\tilde{k}}^t}\ket{\tilde{\jmath},\tilde{k}}\rightarrow \sum_{\tilde{\jmath},\tilde{k}=0}^{N-1} \sqrt{x_{\tilde{\jmath},\tilde{k}}^t}\ket{\tilde{\jmath}\ominus j,\tilde{k}\ominus k}
$$
where $\ominus$ is the difference in base $N$.
\item $\mathcal{A}_T$ as the operator that encodes the template array inside the quantum computer, i.e. $\mathcal{A}_T\ket{0}^{\otimes 2\log N}=\sum_{j,k=0}^{N-1}\sqrt{X_{j,k}^t}\ket{j,k}$, and $\mathcal{A}=\mathcal{A}_T\otimes\mathcal{A}_I\otimes(S_{j,k}\mathcal{A}_I)$.
\item $S_0=\left(I-2\ket{0}\bra{0}\right)^{\otimes 6\log N}_{\text{templC1C2}}$, is the operator reversing the phase of the state $\ket{0}^{\otimes 6 \log N}_{\text{templC1C2}}$, while leaving the other states unchanged. Here, C1 and C2 mean copy 1 and copy 2, respectively.
\item $S_{j''_{\text{C1}}\ominus j'_{\text{templ}}=\bar{\jmath}_{\text{C2}}}=I_{\text{templ, C1, C2}}-2 \sum_{\bar{\jmath},j''=0}^{N-1}\ket{j'',\bar{\jmath}\oplus j'',\bar{\jmath},k'',\bar{k}\oplus k'',\bar{k}}\bra{j'',\bar{\jmath}\oplus j'',\bar{\jmath},k'',\bar{k}\oplus k'',\bar{k}}_{\text{templC1C2}}$, where $\oplus$ is the sum in base $N$, reverse the phase of the states that fulfill the conditions $j''_{\text{C1}}\ominus j'_{\text{templ}}=\bar{\jmath}_{\text{C2}}$ and $k''_{\text{C1}}\ominus k'_{\text{templ}}=\bar{k}_{\text{C2}}$.
\item $G=\mathcal{A} S_0 \mathcal{A}^{-1} S_{j''_{\text{C1}}\ominus j'_{\text{templ}}=\bar{\jmath}_{\text{C2}}}$, known as Grover operator.
\end{enumerate}
For simplicity, we assume the complexity of these operators is 1.
The quantum circuit for the EMML algorithm is in Fig. \ref{Im:EMML}.

\subsection{Input initialization}
\hspace{.5cm}
We apply the gate $\mathcal{A}_T$ to the template set, and the gate $\mathcal{A}_I$ to the two other array sets. The total state after the three gates is
\begin{equation}\begin{split}
&\sum_{j'',k''=0}^{N-1}\sqrt{X_{j'',k''}^t}\ket{j'',k''}_{\text{templ}}\otimes \sum_{j',k'=0}^{N-1}\sqrt{x_{j',k'}^t}\ket{j',k'}_{\text{C1}}\\
&\otimes \sum_{\tilde{\jmath},\tilde{k}=0}^{N-1}\sqrt{x_{\tilde{\jmath},\tilde{k}}^t}\ket{\tilde{\jmath},\tilde{k}}_{\text{C2}}\otimes\ket{0}^{\otimes \log M}_{\text{new}},
\end{split}\end{equation}
where the pedex new stays for new pixel value. We apply the gate $S_{j,k}$ to the third register. The initial state is
\begin{equation}\begin{split}
&\sum_{j'',k''=0}^{N-1}\sqrt{X_{j'',k''}^t}\ket{j'',k''}_{\text{templ}}\otimes \sum_{j',k'=0}^{N-1}\sqrt{x_{j',k'}^t}\ket{j',k'}_{\text{C1}}\\ &\otimes \sum_{\bar{\jmath},\bar{k}=0}^{N-1}\sqrt{x_{j\oplus\bar{\jmath},k\oplus\bar{k}}^t}\ket{\bar{\jmath},\bar{k}}_{\text{C2}}\otimes\ket{0}^{\otimes \log M}_{\text{new}}.\label{Eq:EMMLin}
\end{split}\end{equation}

\subsection{Quantum Amplitude Estimation}
\hspace{.5cm}
We now perform a QAE.
We re-write the state of Eq. \eqref{Eq:EMMLin} in terms of the eigenstates $\ket{\Psi_{\pm}}_{\text{templ, C1, C2}}$ of the Grover operator $G$, i.e.
\begin{equation}
-\frac{i}{\sqrt{2}}\left(e^{i \theta_{j,k}}\ket{\Psi_+}_{\text{templC1C2}}-e^{-i \theta_{j,k}}\ket{\Psi_-}_{\text{templC1C2}}\right),
\end{equation}
where $\sin^2\theta_{j,k}=\sum_{\bar{\jmath},\bar{k}=0}^{N-1}C_{\bar{\jmath},\bar{k}}^t x_{j\oplus\bar{\jmath},k\oplus\bar{k}}^t =x_{j,k}^{t+1}$ is the new pixel value. $\ket{\Psi_{\pm}}_{\text{templC1C2}}$ are defined as
\begin{equation}\begin{split}
\ket{\Psi_{\pm}}&_{\text{templC1C2}}=\\
\frac{1}{\sqrt{2}}&\Biggl(\frac{1}{\sqrt{x_{j,k}^{t+1}}}\sum_{\bar{\jmath},j'=0}^{N-1}\sqrt{X^t_{j'}x^t_{j'\oplus \bar{\jmath}}x^t_{j\oplus \bar{\jmath}}}\ket{j',j'\oplus \bar{\jmath},\bar{\jmath}}_{\text{templC1C2}}\\
+&\frac{1}{\sqrt{1-x_{j,k}^{t+1}}}\sum_{\substack{j',j'',\bar{\jmath}=0 \\ j'\neq \bar{\jmath}\oplus j''}}^{N-1}\sqrt{X^t_{j''}x^t_{j'}x_{j\oplus\bar{\jmath}}}\ket{j'',j',\bar{\jmath}}_{\text{templC1C2}}\Biggl).
\end{split}\end{equation}
The eigenvalues of $\ket{\Psi_{\pm}}$ are $e^{\pm i 2\theta_{j,k}}$. Then, a controlled-$G^M$ $\Gamma_M(G)$ is applied. The operator $\Gamma_M(G)$ is defined as
\begin{equation}
\Gamma_M(G)\ket{y}_{\text{templC1C2}}\ket{m}_{\text{new}}=G^M\ket{y}_{\text{templC1C2}}\ket{m}_{\text{new}}.
\end{equation}
The total state after the controlled-$G^M$ $\Gamma_M (G)$ is
\begin{equation}\begin{split}
-\frac{i}{\sqrt{2}}\bigg(&e^{i\theta_{j,k}}\ket{\Psi_+}_{\text{templC1C2}}\ket{S_M\left(\frac{\theta_{j,k}}{\pi}\right)}_{\text{new}}\\
&-e^{-i\theta_{j,k}}\ket{\Psi_-}_{\text{templC1C2}}\ket{S_M\left(1-\frac{\theta_{j,k}}{\pi}\right)}_{\text{new}}\bigg),
\end{split}\end{equation}
where $\ket{S_M\left(\frac{\theta_{j,k}}{\pi}\right)}_{\text{new}}=\sum_{m=0}^{M-1}e^{2 i m \theta_{j,k}}\ket{m}_{\text{new}}$.
The last QFT applied on the last register renders the state
\begin{equation}\begin{split}
-\frac{i}{\sqrt{2}}\bigg(&e^{i\theta_{j,k}}\ket{\Psi_+}_{\text{templC1C2}}\ket{M\frac{\theta_{j,k}}{\pi}}_{\text{new}}\\
&-e^{-i\theta_{j,k}}\ket{\Psi_-}_{\text{templC1C2}}\ket{M\left(1-\frac{\theta_{j,k}}{\pi}\right)}_{\text{new}}\bigg).
\end{split}\end{equation}
One just has to measure the value $\Theta_{j,k}$ in the register new and calculate $\sin^2 \frac{\pi}{M}\Theta_{j,k}=x_{j,k}^{t+1}$. The average data values $\langle x_{j,k}^{t+1}\rangle$ are of the order of $\frac{1}{N^2}$. Therefore a good choice for $M$ is $M=\alpha N$ with $\alpha\ll N$. The complexity of the algorithm is $O(N)$, that is roughly a quadratical improvement respect to the classical complexity $O(N^2 \log N)$.

{\bf Algorithm for EMML computation.}
\begin{enumerate}
\item Initialize the 3 registers containing the images (template, copy 1, copy 2);
\item Apply a $QFT_M$ to the last register;
\item Apply the operator $\Gamma_M(G)$;
\item Perform an inverse $QFT_M$ on the last register;
\item Measure the new register value $\Theta_{j,k}$;
\item Classically calculate $\sin^2 \frac{\pi}{M}\Theta_{j,k}$.
\end{enumerate}

\section{Concluding remarks}
\hspace{.5cm}
In this article we have explored the possibility of computing cross-correlations using quantum algorithms. We have presented two quantum algorithms both exploiting QAE.\\
The first algorithm consists in computing and storing the cross-correlations in a quantum array. Note that the algorithm can also be used for the computation of the convolution, which is another important tool for signal processing, but also for optical physics, such algorithm opens new possibilities in an even wider range of data analysis than what cross-correlations computation might have done. The algorithm is meant to be only a part of a bigger quantum algorithm. Further studies must focus in investigating what data analysis applications can benefit from it.\\
The second algorithm focus on a specific application of cross-correlations, namely EMML, showing that quantum computing provides roughly a quadratic speed-up for the expectation maximization step. The algorithm can also be extended to other applications such as for convolutional neural networks \cite{LeCun15,LeCun95,Krizhevsky12}.\\
A small remark must be made about the state preparation unitaries used for both algorithms. For the sake of simplicity we have assumed the complexity of these operations to be $O(1)$. Additional studies should focus on how to efficiently perform them. A possibility worth to be investigated is the use of QRAM \cite{Park19}.
Throughout the text the cross-correlation functions have been computed only for real number arrays. This is a limitation that can easily be overcome just noticing how, in the case of complex numbers the cross-correlations are the result of the combination of four real-valued cross-correlations, i.e. the real and imaginary parts of the data arrays. Similarly, EMML algorithms for complex data arrays can be rearranged from the one we presented with only real data arrays noticing that the real and imaginary parts of the new data arrays are given by sums of products of the real and imaginary parts of the initial data arrays.

Concluding, these results, in contrast to previous claims \cite{Lomont03} clearly show that quantum computing can potentially be used in all the algorithms where cross-correlations and convolutions are used.

%

\bibliographystyle{plain}

\end{document}